\documentclass[12pt,preprint,notoc,nohyper]{KSPLB} 

\usepackage{epsfig,multicol,bbm}

\newcommand\fverb{\setbox\pippobox=\hbox\bgroup\verb}
\newcommand\fverbdo{\egroup\medskip\noindent%
            \fbox{\unhbox\pippobox}\ }
\newcommand\fverbit{\egroup\item[\fbox{\unhbox\pippobox}]}
\newbox\pippobox

\title{Tunneling of Dirac particles from accelerating and rotating black holes}

\author{Usman A. Gillani and K. Saifullah  \\

Department of Mathematics, Quaid-i-Azam University, Islamabad,
Pakistan \\

Electronic address: \email{saifullah@qau.edu.pk}}

\preprint{}  

\abstract{Hawking radiation from black holes has been studied as a
phenomenon of quantum tunneling of particles through their horizons.
We have extended this approach to study the tunneling of Dirac
particles from a large class of black holes which includes those
with acceleration and rotation as well. We have calculated the
tunneling probability of incoming and outgoing particles, and
recovered the correct Hawking temperature by this method.}

\dedicated{}

\begin{document}

\section{Introduction}

Hawking's arguments \cite{Hawk1,Hawk2} for thermal radiation and
evaporation of black holes were based on quantum field theoretic
analysis in curved spacetime. This breakthrough gave rise to a whole
new field of investigation where three different areas, the quantum
theory, general relativity and thermodynamics, come together. A
major development in this direction was to interpret Hawking
radiation as a quantum tunneling phenomenon [3-5]. Using this
procedure the Hawking temperature was calculated from the tunneling
probabilities of incoming and outgoing particles \cite{SP, SPS}.
This generated a lot of interest in studying radiations of different
particles in the context of various black hole spacetimes. In this
regard different semiclassical approaches were adopted to study
radiation by scalar and Dirac particles. In particular, a procedure
was set up and tunneling of charged and uncharged Dirac particles
was studied for Kerr, Kerr-Newman, Taub-NUT and G\"{o}del spacetimes
[8-10]. This approach was applied to other four-dimensional [11-15]
and three-dimensional \cite{LR, LLR} spacetimes as well.

In this paper we study a large class [18-23] of Pleba\'{n}ski and
Demia\'{n}ski spacetimes for the tunneling procedure. We are
basically motivated by two reasons. Firstly, the spacetime we study
generalizes many black holes like the Schwarzschild, Kerr and
Taub-NUT, and the tunneling probabilities and Hawking temperatures
for these black holes can be recovered as special cases of our
study. Secondly, the metric we study has very interesting
interpretation in itself as accelerating and rotating black holes
which have been studied in the literature for various properties.
These black holes admit two rotation horizons and two acceleration
horizons. In this paper we apply the procedure mentioned above to
study tunneling of Dirac particles from these black holes. We
calculate their tunneling probability and the Hawking temperature at
these horizons.

\section{Accelerating and rotating black holes}

The metric for accelerating and rotating black holes without the
cosmological constant can be written as \cite{PK, GP05}

\begin{eqnarray}\nonumber
ds^{2} &=&\frac{1}{\Omega ^{2}}\{-(\frac{Q}{\rho
^{2}}-\frac{a^{2}P\sin ^{2}\theta }{\rho ^{2}})dt^{2}+\frac{\rho
^{2}}{Q}dr^{2}+\frac{\rho ^{2}}{P}
d\theta ^{2}  \\
&+&(\frac{P(r^{2}+a^{2})^{2}\sin ^{2}\theta }{\rho
^{2}}-\frac{Qa^{2}\sin ^{4}\theta }{\rho ^{2}})\}d\phi
^{2}-\frac{2a\sin ^{2}\theta(P(r^{2}+a^{2})-Q)dtd\phi }{\rho
^{2}\Omega ^{2}}.
\end{eqnarray}
where
\begin{eqnarray}
\Omega  &=&1-\alpha r\cos \theta , \\
\rho ^{2} &=&r^{2}+a^{2}\cos ^{2}\theta , \\
P &=&1-2\alpha M\cos \theta +\alpha ^{2}a^{2}\cos ^{2}\theta , \\
Q &=&(a^{2}-2Mr+r^{2})(1-\alpha ^{2}r^{2}).  \label{1.56}
\end{eqnarray}
Here the parameters $M$, $a$ and $\alpha $ represent the mass,
angular momentum per unit mass, and acceleration of the source
respectively. Following the notation of Ref. \cite{KM08a} the above
metric can be written as
\begin{equation}
ds^{2}=-f(r,\theta )dt^{2}+\frac{dr^{2}}{g(r,\theta )}+\Sigma
(r,\theta )d\theta ^{2}+K(r,\theta )d\phi ^{2}-2H(r,\theta )dtd\phi
,
\end{equation}
where $f(r,\theta ), {g(r,\theta )}, \Sigma (r,\theta ), K(r,\theta
),H(r,\theta )$ are defined below
\begin{eqnarray}
f(r,\theta ) &=&\frac{1}{\Omega ^{2}}\{\frac{Q}{\rho ^{2}}-\frac{
a^{2}P\sin ^{2}\theta }{\rho ^{2}}\}, \label{9} \\
g(r,\theta ) &=&\frac{Q\Omega ^{2}}{\rho ^{2}}, \label{10} \\
\Sigma (r,\theta ) &=&\frac{\rho ^{2}}{P\Omega ^{2}}, \label{11} \\
K(r,\theta ) &=&\frac{P(r^{2}+a^{2})^{2}\sin ^{2}\theta }{\rho
^{2}\Omega
^{2}}-\frac{Qa^{2}\sin ^{4}\theta }{\rho ^{2}\Omega ^{2}}, \label{12} \\
H(r,\theta ) &=&\frac{a\sin ^{2}\theta(P(r^{2}+a^{2})-Q)}{\rho
^{2}\Omega ^{2}}. \label{13}
\end{eqnarray}
The event horizons of this black hole can be calculated by putting
\begin{equation}
\frac{1}{g_{11}}=0,
\end{equation}
which implies that
\begin{equation}
g(r,\theta )=\frac{Q\Omega ^{2}}{\rho ^{2}}=0,
\end{equation}
or
\begin{equation}
\Omega=0, Q=0,
\end{equation}
which shows that in addition to the outer and inner horizons
corresponding to the Kerr-Newman black hole horizons

\begin{equation}
r_{\pm}=M\pm \sqrt{M^{2}-a^{2}}, \label{222}
\end{equation}
we get two acceleration horizons \cite{GP05, GP06}:
$r=\frac{1}{\alpha}$ and $r=\frac{1}{\alpha \cos \theta}$.

We also define the function \cite{KM08a}
\begin{equation}
F(r,\theta )=f(r,\theta )+\frac{H^{2}(r,\theta )}{K(r,\theta )}.
\end{equation}
Putting the values of $f(r,\theta ),K(r,\theta ), H(r,\theta )$ in
this we get
\begin{equation}
F(r,\theta )=\frac{QP\rho ^{2}}{(P(r^{2}+a^{2})^{2}-Qa^{2}\sin
^{2}\theta) \Omega ^{2}}. \label{21}
\end{equation}
The angular velocity, for the above metric takes the form
\cite{KM08a}
\begin{equation}
\Omega _{H}=\frac{H(r_{+},\theta )}{K(r_{+},\theta )}. \label{22}
\end{equation}
Substituting $H(r_{+},\theta ),K(r_{+},\theta)$ we get
\begin{equation}
\Omega _{H}=\frac{a(P(r_{+}^{2}+a^{2})-Q(r_{+})}{Q(r_{+})a^{2}\sin
^{2}\theta +P(r_{+}^{2}+a^{2})^{2}} . \label{23}
\end{equation}
Using the fact that $Q(r_{+})=0$, this can be written as
\begin{equation}
\Omega _{H}=\frac{a}{r_{+}^{2}+a^{2}}. \label{24}
\end{equation}

\section{Tunneling of Dirac particles}

In order to study the tunneling of Dirac particles we solve the
Dirac equation in the back ground of accelerating and rotating black
holes. The covariant Dirac equation can be written as
\begin{equation}
\iota \gamma ^{\mu }(D_{\mu })\Psi +\frac{m}{\hbar }\Psi =0,
\end{equation}
where we have
\begin{equation}
D_{\mu }=\partial _{\mu }+\Omega _{\mu },
\end{equation}

\begin{equation}
\Omega _{\mu }=\frac{-1}{8}\Gamma _{\mu }^{\alpha \beta }[\gamma
^{\alpha },\gamma ^{\beta }],
\end{equation}
and $[\gamma ^{\alpha },\gamma ^{\beta }]$ satisfy the commutation
relations
\begin{equation}
\lbrack \gamma ^{\alpha },\gamma ^{\beta }]=-[\gamma ^{\beta
},\gamma ^{\alpha }] \hspace{10pt} \textrm{if} \hspace{10pt} \alpha
\neq \beta , \hspace{6pt} [\gamma ^{\alpha },\gamma ^{\beta }]=0
\hspace{10pt} \textrm{if} \hspace{10pt} \alpha =\beta . \label{100}
\end{equation}
Here $m$ is the mass of the fermions and $\gamma ^{\mu }$ matrices
satisfy $[{\gamma ^{\alpha },\gamma ^{\beta }}]=2g^{\mu \nu }I$,
($I$ is the identity matrix). For fermion tunneling radiation, it is
important to choose appropriate $\gamma ^{\mu }$ matrices. We take

\begin{equation}
\gamma ^{t}=\sqrt{\frac{(P(r^{2}+a^{2})^{2}-Qa^{2}\sin
^{2}\theta)\Omega ^{2}}{ PQ\rho ^{2}}}\gamma ^{0}, \gamma
^{r}=\sqrt{\frac{Q\Omega ^{2}}{\rho ^{2}}}\gamma ^{3}, \gamma
^{\theta }=\sqrt{ \frac{P\Omega ^{2}}{\rho ^{2}}}\gamma ^{1},
\end{equation}
\begin{equation}
\gamma ^{\phi }=\frac{\rho \Omega \gamma ^{2}}{\sin \theta
\sqrt{P(r^{2}+a^{2})^{2}-Qa^{2}\sin ^{2}\theta }}+\frac{a
(P(r^{2}+a^{2})-Q)\gamma
^{0}}{\sqrt{F(r,\theta)}(P(r^{2}+a^{2})^{2}-Qa^{2}\sin ^{2}\theta
)}.
\end{equation}
Here
\begin{equation}
\gamma ^{0}=\left(
\begin{array}{cc}
0 & I \\
-I & 0 \\
&
\end{array}
\right) , \gamma ^{1}=\left(
\begin{array}{cc}
0 & \sigma ^{1} \\
\sigma ^{1} & 0 \\
&
\end{array}
\right) ,  \end{equation}
\begin{equation}
\gamma ^{2}=\left(
\begin{array}{cc}
0 & \sigma ^{2} \\
\sigma ^{2} & 0 \\
&
\end{array}
\right) , \gamma ^{3}=\left(
\begin{array}{cc}
0 & \sigma ^{3} \\
\sigma ^{3} & 0 \\
&
\end{array}
\right) .
\end{equation}
The $\sigma ^{i}(i =1,2,3)$ are the Pauli sigma matrices given by

\begin{equation}
\sigma ^{1}=\left(
\begin{array}{cc}
0 & 1 \\
1 & 0 \\
&
\end{array}
\right) ,\sigma ^{2}=\left(
\begin{array}{cc}
0 & -\iota  \\
\iota  & 0 \\
&
\end{array}
\right) ,  \end{equation}
\begin{equation}
\sigma ^{3}=\left(
\begin{array}{cc}
1 & 0 \\
0 & -1 \\
&
\end{array}
\right) .
\end{equation}
In order to solve the Dirac equation we assume the following ansatz
involving arbitrary functions of the coordinates corresponding to
the spin-up and spin-down solutions of this equation

\begin{eqnarray}
\Psi_{\uparrow}(t,r,\theta ,\phi )=\left(
\begin{array}{c}
A(t,r,\theta ,\phi )\xi_{\uparrow} \\
B(t,r,\theta ,\phi )\xi_{\uparrow} \\
\end{array}
\right) \exp[\frac{\iota I_{\uparrow}(t,r,\theta ,\phi )}{\hbar }],
 \label{101} \\
\Psi_{\downarrow}(t,r,\theta ,\phi )=\left(
\begin{array}{c}
C(t,r,\theta ,\phi )\xi _{\downarrow } \\
D(t,r,\theta ,\phi )\xi _{\downarrow } \\
\end{array}
\right) \exp[\frac{\iota I_{\downarrow }(t,r,\theta ,\phi )}{\hbar
}]. \label{102}
\end{eqnarray}
Here $\xi _{\uparrow}$ and $\xi _{\downarrow}$ are the eigenvectors
of  $\sigma ^{3}$, and $I_{\uparrow}$ and $I_{\downarrow}$ denote
the actions of the emitted spin-up and spin-down particles,
respectively. On using Eq. (\ref{100}) the Dirac equation takes the
form
\begin{equation}
(\iota \gamma ^{t}\partial _{t}+\iota \gamma ^{r}\partial _{r}+\iota
\gamma ^{\theta }\partial _{\theta }+\iota \gamma ^{\phi }\partial
_{\phi })\Psi + \frac{m}{\hbar }\Psi =0.
\end{equation}
Now, we substitute the above ansatz into the Dirac equation and
compute it term by term. Dividing by the exponential term and
neglecting the terms with $\hbar$ we obtain the following four
equations

\begin{eqnarray}\nonumber
0 &=&-B[\frac{1}{\sqrt{F(r,\theta )}}\partial _{t}I_{\uparrow
}+\sqrt{\frac{ \Omega ^{2}Q}{\rho ^{2}}}\partial
_{r}I_{\uparrow }   \\
 &+&\frac{a(P(r^{2}+a^{2})-Q)}{\sqrt{F(r,\theta )}(P(r^{2}+a^{2})^{2}-Qa^{2}\sin ^{2}\theta )}\partial _{\phi}I_{\uparrow }]+Am, \\
0 &=&-B[\sqrt{\frac{\Omega ^{2}P}{\rho ^{2}}}\partial _{\theta
}I_{\uparrow }+\frac{\iota \rho \Omega }{\sin \theta
(\sqrt{P(r^{2}+a^{2})^{2}-Qa^{2}\sin ^{2}\theta )}}\partial _{\phi
}I_{\uparrow }], \\ \nonumber 0&=&A[\frac{1}{\sqrt{F(r,\theta
)}}\partial _{t}I_{\uparrow }-\sqrt{\frac{
\Omega ^{2}Q }{\rho ^{2}}}\partial _{r}I_{\uparrow }  \\
&+&\frac{a(P(r^{2}+a^{2})-Q)}{\sqrt{F(r,\theta )}(P(r^{2}+a^{2})^{2}-Qa^{2}\sin ^{2}\theta )}\partial _{\phi}I_{\uparrow }]+Bm, \\
0 &=&A[\sqrt{\frac{\Omega ^{2}P}{\rho ^{2}}}\partial _{\theta
}I_{\uparrow }+\frac{\iota \rho \Omega }{\sin \theta
\sqrt{P(r^{2}+a^{2})^{2}-Qa^{2}\sin ^{2}\theta }}\partial _{\phi
}I_{\uparrow }].
\end{eqnarray}
Taking into account the symmetries of the spacetime we assume the
action to be of the form

\begin{equation}
I_{\uparrow }=-Et+J\phi +W(r,\theta ).
\end{equation}
Here $E$ and $J$ denote the energy and angular momentum of the
emitted particle. We do calculations for the spin-up case; the
spin-down case is similar. Inserting this into the above four
equations, we get

\begin{eqnarray}\nonumber
0 &=&-B[\frac{-E}{\sqrt{F(r,\theta )}}+\sqrt{\frac{\Omega ^{2}Q }{
\rho ^{2}}}W^{\prime }(r,\theta )  \label{50} \\
&+&\frac{a(P(r^{2}+a^{2})-Q)}{\sqrt{F(r,\theta )}(P(r^{2}+a^{2})^{2}-Qa^{2}\sin ^{2}\theta )}J]+Am, \\
0 &=&B[\sqrt{\frac{\Omega ^{2}P}{\rho ^{2}}}W_{\theta }(r,\theta
)+\frac{ \iota \rho \Omega }{\sin \theta
(\sqrt{P(r^{2}+a^{2})^{2}-Qa^{2}\sin ^{2}\theta }}J], \label{51} \\
\nonumber 0&=&A[\frac{-E}{\sqrt{F(r,\theta )}}-\sqrt{\frac{\Omega
^{2}Q }{
\rho ^{2}}}W^{\prime }(r,\theta )  \\
&+&\frac{a(P(r^{2}+a^{2})-Q)}{\sqrt{F(r,\theta )}(P(r^{2}+a^{2})^{2}-Qa^{2}\sin ^{2}\theta )}J]+B m, \label{52} \\
0 &=&A[\sqrt{\frac{\Omega ^{2}P}{\rho ^{2}}}W_{\theta }(r,\theta
)+\frac{ \iota \rho \Omega }{\sin \theta
\sqrt{P(r^{2}+a^{2})^{2}-Qa^{2}\sin ^{2}\theta }}J]. \label{53}
\end{eqnarray}
Expanding Eq. (\ref{10}) in Taylor's series and neglecting the
higher powers we get

\begin{equation}
g(r,\theta )=g(r_{+},\theta )+(r-r_{+})g_{r}(r_{+},\theta ).
\end{equation}
Noting that at the horizon $g(r_{+},\theta )=0$, and its derivative
at the horizon is

\begin{equation}
g_{r}(r_{+},\theta )=\frac{2(1-\alpha r_{+} \cos \theta
)^{2}(r_{+}-M)(1-\alpha ^{2}r_{+}^{2})}{r_{+}^{2}+a^{2}\cos
^{2}\theta },
\end{equation}
we obtain

\begin{equation}
g(r,\theta )=(r-r_{+})\frac {2(1-\alpha r_{+} \cos \theta
)^{2}(r_{+}-M)(1-\alpha ^{2}r_{+}^{2})}{r_{+}^{2}+a^{2}\cos
^{2}\theta}. \label{58}
\end{equation}
Similarly, noting that at the horizon $F(r_{+},\theta)=0$, from Eq.
(\ref{21}) we obtain

\begin{equation}
F(r,\theta )=(r-r_{+})(\frac{2(r_{+}^{2}+a^{2}\cos ^{2}\theta
)(r_{+}-M)(1-\alpha ^{2}r_{+}^{2})}{(r_{+}^{2}+a^{2})^{2}\Omega
^{2}} ). \label{59}
\end{equation}
Now expanding Eqs. (\ref{50}) - (\ref{53}) near the black hole
horizon and using Eqs. (\ref{58}) and (\ref{59}) we get

\begin{eqnarray}\nonumber
0 &=&-B[\frac{-E}{\sqrt{(r-r_{+})F_{r}(r_{+},\theta )}}+\sqrt{
(r-r_{+})g_{r}(r_{+},\theta )}W^{\prime }(r,\theta )  \\
&+&\frac{a(P(r_{+}^{2}+a^{2})-Q(r_{+}))}{\sqrt{(r-r_{+})F_{r}(r_{+},\theta
)}
(P(r_{+}^{2}+a^{2})^{2}-Q(r_{+})a^{2}\sin ^{2}\theta )}J]+Am, \label{60} \\
0 &=&-B[\sqrt{\frac{\Omega ^{2}(r_{+},\theta )P}{\rho
^{2}(r_{+},\theta )}}W_{\theta }(r,\theta )+\frac{ \iota \rho(r_+,
\theta) \Omega (r_+, \theta)}{\sin \theta
\sqrt{P(r_{+}^{2}+a^{2})^{2}-Q(r_+)a^{2}\sin ^{2}\theta }}J], \label{61} \\
\nonumber 0&=&A[\frac{-E}{\sqrt{(r-r_{+})F_{r}(r_{+},\theta
)}}-\sqrt{(r-r_+)g_r(r_+, \theta)}W^{\prime }(r,\theta ),  \\
&+&\frac{a(P(r_{+}^{2}+a^{2})-Q(r_{+}))}{\sqrt{(r-r_{+})F_{r}(r_{+},\theta
)}
(P(r_{+}^{2}+a^{2})^{2}-Q(r_{+})a^{2}\sin ^{2}\theta )}J]+Bm, \label{62} \\
0 &=&-A[\sqrt{\frac{\Omega ^{2}(r_{+},\theta )P}{\rho
^{2}(r_{+},\theta )}}W_{\theta }(r,\theta )+\frac{ \iota \rho(r_+,
\theta) \Omega (r_+, \theta)}{\sin \theta
\sqrt{P(r_{+}^{2}+a^{2})^{2}-Q(r_+)a^{2}\sin ^{2}\theta }}J].
\label{63}
\end{eqnarray}
We neglect the equations that depend upon $``\theta''$, as their
contribution to the total tunneling rate is canceled out, and retain
only the radial equations (Eqs. (\ref{60}) and (\ref{62})) which on
using Eq. (\ref{23}) take the form

\begin{eqnarray}
0 &=&-B[\frac{-E+\Omega _{H}J}{\sqrt{(r-r_{+})F_{r}(r_{+},\theta
)}}+\sqrt{ (r-r_{+})g_{r}(r_{+},\theta )}W^{\prime }(r,\theta )]
+Am, \label{64} \\
 0&=&A[\frac{-E+\Omega
_{H}J}{\sqrt{(r-r_{+})F_{r}(r_{+},\theta )}}-\sqrt{
(r-r_{+})g_{r}(r_{+},\theta )}W^{\prime }(r,\theta )]+Bm. \label{65}
\end{eqnarray}
At the horizon we can further seperate $W(r,\theta )$ as
\begin{equation}
W(r,\theta )=W(r)+\Theta (\theta )
\end{equation}
In the massless case there exist two possible solutions
\begin{equation} \nonumber
B=0,W^{\prime }(r)=W_{+}^{\prime }(r)=\frac{(E-\Omega _{H}J)}{
\sqrt{(r-r_{+})F_{r}(r_{+},\theta
)}\sqrt{(r-r_{+})g_{r}(r_{+},\theta )}},
\end{equation}
and
\begin{equation}
A=0,W^{\prime }(r)=W_{-}^{\prime }(r)=\frac{(-E+\Omega _{H}J)}{
\sqrt{(r-r_{+})F_{r}(r_{+},\theta
)}\sqrt{(r-r_{+})g_{r}(r_{+},\theta )}}.
\end{equation}
Substituting the values of $F_{r}(r_{+},\theta )$ and
$g_{r}(r_{+},\theta )$ the above equations become
\begin{eqnarray}
W_{+}^{\prime }(r) &=&\frac{(E-\Omega _{H}J)(r_{+}^{2}+a^{2})}{
2(r-r_{+})(r_{+}-M)(1-\alpha ^{2}r_{+}^{2})},  \\
 W_{-}^{\prime }(r) &=&\frac{(-E+\Omega _{H}J)(r_{+}^{2}+a^{2})}{
2(r-r_{+})(r_{+}-M)(1-\alpha ^{2}r_{+}^{2})}.
\end{eqnarray}
Here prime denotes the derivative with respect to $r$ and $+/-$
correspond to the outgoing/incoming solutions. For finding the value
of $W(r)$ we integrate the above result
\begin{eqnarray}
W_{+}(r) &=&\int \frac{(E-\Omega _{H}J)(r_{+}^{2}+a^{2})}{
2(r-r_{+})(r_{+}-M)(1-\alpha ^{2}r_{+}^{2})}.
\end{eqnarray}
Integrating around the pole $r=r_{+}$ this gives
\begin{equation}
W_{+}(r)=\pi \iota \frac{(E-\Omega _{H}J)(r_{+}^{2}+a^{2})}{
2(r_{+}-M)(1-\alpha ^{2}r_{+}^{2})}.
\end{equation}
Dropping the subscript we write
\begin{eqnarray}
W(r) &=&\pi \iota \frac{(E-\Omega _{H}J)(r_{+}^{2}+a^{2})}{
2(r_{+}-M)(1-\alpha ^{2}r_{+}^{2})},  \\
ImW &=&\pi \frac{(E-\Omega
_{H}J)(r_{+}^{2}+a^{2})}{2(r_{+}-M)(1-\alpha ^{2}r_{+}^{2})}.
\end{eqnarray}
So the tunneling probabilities of fermions are
\begin{eqnarray}
P_{emission}= \exp[-2ImI] &=&\exp[-2(ImW_{+}+Im\Theta )], \label{a} \\
P_{absorption}=  \exp[-2ImI] &=&\exp[-2(ImW_{-}+Im\Theta )].
\end{eqnarray}
Since $ImW_{+}=-ImW_{-}$, we see that the total probability that the
particle tunnels from inside the event horizon to outside is
\begin{equation}
\Gamma \sim \frac{P_{emission}}{P_{absorption}} = \exp[-4ImW_{+}],
\end{equation}
or
\begin{equation}
\Gamma =\exp[\frac{-2\pi (E-\Omega _{H}J)(r_{+}^{2}+a^{2})}{
(r_{+}-M)(1-\alpha ^{2}r_{+}^2)}]. \label{b}
\end{equation}
Comparing this with $\Gamma=\exp [-\beta E]$ where $\beta = 1/T_H$
we find that the Hawking temperature \cite{SP, SPS} is given by

\begin{equation}
T_{H}=\frac{\left( r_{+}-M\right) \left( 1-\alpha
^{2}r_{+}^{2}\right) }{2\pi \left( r_{+}^{2}+a^{2}\right)} ,
\label{c}
\end{equation}
where $r_+$ is given by Eq. (\ref{222}). If we put acceleration
equal to zero in formulae (\ref{b}) and (\ref{c}), they reduce to
the tunneling probability and temperature of the Kerr black hole
\cite{KM06, BS}. Similarly, setting rotation equal to zero will
recover expressions for the Schwarzschild black hole. Comparing with
the Kerr black hole, we note that the effect of acceleration is that
it increases the temperature.

From Eqs. (\ref{a})-(\ref{b}) it appears that for some value of $E$
and $J$ the probabilities may become larger than 1 and hence violate
unitarity. However, this is not the case because apart from the
spatial contribution there is also a contribution to the imaginary
part $Im(E\Delta t)$, from the temporal part of the action
\cite{APGS, APS}. This shifts the time by an imaginary amount which
contributes both to $P_{emission}$ and $P_{absorption}$ and results
in a correct value of $\Gamma$. If we do not take this contribution
into account, we will obtain the Hawking temperature twice as large
as the actual value \cite{AAS, Pi}. Further, we note that
$P_{absorption}$ will actually be equal to 1, because the
trajectories of incoming particles do not face any barrier. This is,
in fact, taken care of if the temporal contribution is also taken
into account \cite{APGS, APS}, and we obtain the correct value of
the tunneling probability.

For the massive case Eqs. (\ref{60}) and (\ref{62}) do not decouple.
We eliminate the function $ W^{\prime }(r,\theta )$ from these two
equations by multiplying Eq. (\ref{62}) by $B$ and Eq. (\ref{60}) by
$A$ and subtracting to yield

\begin{eqnarray}
\frac{A}{B} &=&\frac{-(E-J\Omega _{H})\pm \sqrt{(E-J\Omega
_{H})^{2}+m^{2}F_{r}(r_{+},\theta
)(r-r_{+})}}{m\sqrt{F_{r}(r_{+},\theta )(r-r_{+})}}.
\end{eqnarray}
In the limit $r\rightarrow r_{+}$ the two roots give either
$\frac{A}{B}\rightarrow 0$ or $\frac{A}{B} \rightarrow -\infty $,
i.e. either $A\rightarrow 0$ or $B\rightarrow 0$. For $A\rightarrow
0$ we find the value of $m$ from Eq. (\ref{65})
\begin{equation}
m=\frac{-A}{B}(\frac{-E+J\Omega _{H}}{\sqrt{F_{r}(r_{+},\theta
)(r-r_{+})}}- \sqrt{g_{r}(r_{+},\theta )(r-r_{+})}W^{\prime }(r)).
\end{equation}
Putting in Eq. (\ref{64}) and simplifying we get
\begin{equation}
W_{r}(r,\theta )=W_{+}^{\prime }(r)=\frac{(E-J\Omega _{H})(1+ A^{2}/
B^{2})}{\sqrt{F_{r}(r_{+},\theta )g_{r}(r_{+},\theta )}(r-r_{+})(1-
A^{2}/B^{2})}.
\end{equation}
Integrating with respect to $r$ as done before we finally get

\begin{eqnarray}
W(r) &=&\pi \iota \frac{(E-\Omega _{H}J)(r_{+}^{2}+a^{2})}{
2(r_{+}-M)(1-\alpha ^{2}r_{+}^{2})},  \\
ImW &=&\pi \frac{(E-\Omega
_{H}J)(r_{+}^{2}+a^{2})}{2(r_{+}-M)(1-\alpha ^{2}r_{+}^{2})}.
\end{eqnarray}
For $B\rightarrow 0$ we simply get

\begin{equation}
W_{r}(r,\theta )= W_{-}(r)=\pi \iota \frac{-(E-\Omega
_{H}J)(r_{+}^{2}+a^{2})}{ 2(r_{+}-M)(1-\alpha ^{2}r_{+}^{2})}.
\end{equation}
We note that we obtain the same tunneling probabilities as before,
and hence the same temperature. This is because near the black hole
horizon massive particles behave as massless.

\section{The acceleration horizon}

As mentioned earlier the black holes under consideration have an
acceleration horizon at $r_{\alpha}=1/\alpha$ also, apart from the
rotation horizons. The functions $F_{r}(r,\theta )$ and $
g_{r}(r,\theta )$ in this case will be
\begin{eqnarray}
F_{r}(r_{\alpha},\theta ) &=&\frac{(r_{\alpha}^{2}+a^{2}\cos
^{2}\theta )(a^{2}-2Mr_{\alpha}+r_{\alpha}^{2})(-2r_{\alpha}\alpha
^{2})}{(r_{\alpha}^{2}+a^{2})^{2}
\Omega ^{2}(r_{\alpha}, \theta )},   \\
g_{r}(r_{\alpha},\theta ) &=&\frac{(1-\cos \theta )^{2}(\alpha
^{2}a^{2}-2M\alpha +1)(-2\alpha )}{(1+\alpha ^{2}a^{2}\cos
^{2}\theta )},
\end{eqnarray}
and from Eqs. (\ref{64}) and (\ref{65}) we see that the massless
case gives rise to two possible solutions
\begin{equation}
B=0, W^{\prime }(r)=W_{+}^{\prime }(r)=\frac{-(-E+\Omega _{H}J)}{
\sqrt{(r-r_{\alpha})F_{r}(r_{\alpha},\theta
)}\sqrt{(r-r_{\alpha})g_{r}(r_{\alpha},\theta )}}.
\end{equation}
and

\begin{equation}
A=0, W^{\prime }(r)=W_{-}^{\prime }(r)=\frac{(-E+\Omega _{H}J)}{
\sqrt{(r-r_{\alpha})F_{r}(r_{\alpha},\theta
)}\sqrt{(r-r_{\alpha})g_{r}(r_{\alpha},\theta )}}.
\end{equation}
Putting values of the functions $F_{r}(r_{+},\theta )$ and $
g_{r}(r_{+},\theta )$ in the above equations and integrating as
before we obtain
\begin{eqnarray}
W_{+}(r)&=&\frac{\pi \iota (E-\Omega _{H}J)(1+\alpha
^{2}a^{2})}{2\alpha \left( \alpha ^{2}a^{2}-2M\alpha +1\right) }.
\end{eqnarray}
Similarly
\begin{equation}
W_{-}(r)=\frac{-\pi \iota (E-\Omega _{H}J)(1+\alpha
^{2}a^{2})}{2\alpha \left( \alpha ^{2}a^{2}-2M\alpha +1\right) }.
\end{equation}
Proceeding as before to find the tunneling probability, $\Gamma=\exp
[-\beta E]$, the resulting Hawking temperature at the acceleration
horizon comes out to be

\begin{equation}
T_{H}=\frac{\alpha \left( \alpha ^{2}a^{2}-2M\alpha +1\right) }{2\pi
(1+\alpha ^{2}a^{2}) }.
\end{equation}

\section{Calculation of the action}

Here we will work out the action explicitly. We have already seen
that $r$- and $\theta$-dependence decouples in Eqs. (\ref{60}) -
(\ref{63}) near the black hole horizon. This allows us to separate
the action as
\begin{equation}
W(r,\theta )=R(r)+\Theta (\theta ).
\end{equation}
Using this separation in Eqs. (\ref{64}) and (\ref{65}) we get
\begin{eqnarray}
0 &=&-B(\frac{-E+\Omega _{H}J}{\sqrt{(r-r_{+})F_{r}(r_{+},\theta
)}}+\sqrt{ (r-r_{+})g_{r}(r_{+},\theta )}R^{\prime }(r))+Am,
\label{108} \\  0&=&A(\frac{-E+\Omega
_{H}J}{\sqrt{(r-r_{+})F_{r}(r_{+},\theta )}}-\sqrt{
(r-r_{+})g_{r}(r_{+},\theta )}R^{\prime }(r))+Bm. \label{109}
\end{eqnarray}
If $m=0$, from Eq. (\ref{108}) we get
\begin{equation}
B=0,  R^{\prime }(r)=R_{+}^{\prime }(r)=\frac{E-\Omega _{H}J}{
\sqrt{(r-r_{+})F_{r}(r_{+},\theta
)}\sqrt{(r-r_{+})g_{r}(r_{+},\theta )}}.
\end{equation}
Similarly from Eq. (\ref{109}) we get
\begin{equation}
A=0, R^{\prime }(r)=R_{-}^{\prime }(r)=\frac{-E+\Omega _{H}J}{
\sqrt{(r-r_{+})F_{r}(r_{+},\theta
)}\sqrt{(r-r_{+})g_{r}(r_{+},\theta )}}.
\end{equation}
Substituting the values of $F_{r}(r_{+},\theta )$ and $
g_{r}(r_{+},\theta )$ in these equations gives

\begin{eqnarray}
R_{+}^{\prime }(r) &=&\frac{(E-\Omega _{H}J)(r_{+}^{2}+a^{2})}{
2(r-r_{+})(r_{+}-M)(1-\alpha ^{2}r_{+}^{2})},  \label{112}  \\
R_{-}^{\prime }(r) &=&\frac{(-E+\Omega _{H}J)(r_{+}^{2}+a^{2})}{
2(r-r_{+})(r_{+}-M)(1-\alpha ^{2}r_{+}^{2})}, \label{113}
\end{eqnarray}
Integrating Eqs. (\ref{112}) and (\ref{113}) we get
\begin{eqnarray}
R_{+}(r) &=&\frac{(E-\Omega
_{H}J)(r_{+}^{2}+a^{2})}{2(r_{+}-M)(1-\alpha
^{2}r_{+}^{2})}\ln(r-r_{+}).
\end{eqnarray}
\begin{equation}
R_{-}(r)=\frac{-(E-\Omega
_{H}J)(r_{+}^{2}+a^{2})}{2(r_{+}-M)(1-\alpha
^{2}r_{+}^{2})}\ln(r-r_{+}).
\end{equation}
For the massive case from Eq. (\ref{108}) we get
\begin{equation}
R^{\prime }\left( r\right) =\frac{Am}{B\sqrt{\left( r-r_{+}\right)
g_{r}\left( r_{+},\theta \right) }}+\frac{E-\Omega _{H}J}{\left(
r-r_{+}\right) \sqrt{g_{r}\left( r_{+},\theta \right) F_{r}\left(
r_{+}\theta \right) }},
\end{equation}
where $A$ and $B$ are functions of $\left( t,r,\theta ,\phi
\right)$. After integrating it with respect to $r$ and using the
values of $g_{r}\left(r_{+},\theta \right) $ and $F_{r}\left( r_{+},
\theta \right)$ we get for the outgoing particle
\begin{equation}
R_{+}\left( r\right) =\int \frac{Am}{B\sqrt{\left( r-r_{+}\right)
g_{r}\left( r_{+},\theta \right) }}dr+\frac{\left( E-\Omega
_{H}J\right) \left( r_{+}^{2}+a^{2}\right) }{2 \left( r_{+}-M\right)
\left( 1-\alpha ^{2}r_{+}^{2}\right) }\ln \left( r-r_{+}\right) .
\end{equation}
Similarly from Eq. (\ref{109}), for the incoming particle we get
\begin{equation}
R_{-}\left( r\right) =\int \frac{Bm}{A\sqrt{\left( r-r_{+}\right)
g_{r}\left( r_{+},\theta \right) }}dr-\frac{\left( E-\Omega
_{H}J\right) \left( r_{+}^{2}+a^{2}\right) }{2 \left( r_{+}-M\right)
\left( 1-\alpha ^{2}r_{+}^{2}\right) }\ln \left( r-r_{+}\right) .
\end{equation}

Now, we come to Eqs. (\ref{61}) and (\ref{63}) which take the form
\begin{eqnarray}
0 &=&-B(\sqrt{\frac{\Omega ^{2}P}{\rho ^{2}}}\Theta _{\theta }+
\frac{\iota \rho \Omega }{\sin \theta (\sqrt{P(r^{2}+a^{2})^{2}-Qa^{2}\sin ^{2}\theta }}J),  \\
0 &=&-A(\sqrt{\frac{\Omega ^{2}P}{\rho ^{2}}}\Theta _{\theta }+
\frac{\iota \rho \Omega }{\sin \theta
(\sqrt{P(r^{2}+a^{2})^{2}-Qa^{2}\sin ^{2}\theta )^{2}}}J).
\end{eqnarray}
Note that both the equations give

\begin{equation}
\Theta _{\theta }=\frac{-\iota \rho ^{2}J}{\sin \theta
\sqrt{P}\sqrt{ P(r^{2}+a^{2})^{2}-Qa^{2}\sin ^{2}\theta }},
\end{equation}
or at horizon

\begin{equation}
\Theta _{\theta }=\frac{-\iota \rho ^{2}(r_{+},\theta )J}{
(r_{+}^{2}+a^{2})P\sin \theta }.
\end{equation}
Substituting the values of $\rho$ and $P$, and integrating we get
\begin{equation}
\Theta (\theta )=\frac{-\iota J}{(r_{+}^{2}+a^{2})}\int \frac{
(r_{+}^{2}+a^{2}\cos ^{2}\theta )d\theta }{\sin \theta \lbrack
1-2\alpha M\cos \theta +\alpha ^{2}a^{2}\cos ^{2}\theta ]}.
\end{equation}
Using the method of partial fractions we evaluate the integral and
finally get

\begin{eqnarray*}\nonumber
\Theta \left( \theta \right) &=&L_{1}\ln (1+\cos \theta )-L_{2}\ln
(1-\cos \theta )+L_{3}\ln (1-2\alpha M\cos \theta +\alpha
^{2}a^{2}\cos ^{2}\theta )
\\
&&+L_{4}\ln \left( \frac{\alpha a^{2}\cos \theta -M-\sqrt{M^{2}-a^{2}}}{%
\alpha a^{2}\cos \theta -M+\sqrt{M^{2}-a^{2}}}\right) -L_{5}\ln \left( \frac{%
\alpha a^{2}\cos \theta -M-\sqrt{M^{2}-a^{2}}}{\alpha a^{2}\cos \theta -M+%
\sqrt{M^{2}-a^{2}}}\right) .
\end{eqnarray*}
where $L_i$ are

\begin{eqnarray*}
L_{1} &=&\frac{iJ}{2\left( 1+2\alpha M+\alpha ^{2}a^{2}\right) }, \\
L_{2} &=&\frac{iJ}{2\left( 1-2\alpha M+\alpha ^{2}a^{2}\right) }, \\
L_{3} &=&\frac{iJ\alpha M}{(1-2\alpha M+\alpha ^{2}a^{2})(1+2\alpha
M+\alpha
^{2}a^{2})}, \\
L_{4} &=&\frac{iJ\alpha ^{2}\left[ a^{2}\left( 1+\alpha
^{2}a^{2}\right) -2M^{2}\right] }{2\alpha
\sqrt{M^{2}-a^{2}}(1-2\alpha M+\alpha
^{2}a^{2})(1+2\alpha M+\alpha ^{2}a^{2})}, \\
L_{5} &=&\frac{iJa^{2}}{2\alpha \sqrt{M^{2}-a^{2}}\left(
r_{+}^{2}+a^{2}\right) }.
\end{eqnarray*}
This determines the action completely.

\section{Conclusion}

For studying Hawking radiations from black holes different types of
approaches have been adopted in the literature. This has been done
by using the Newman-Penrose formalism and the so-called
Hamilton-Jacobi method. In particular, tunneling of Dirac particles
has been studied for the Kerr and Schwarzschild black holes. We have
extended this semi-classical approach to study a large class of
black holes which include those with acceleration and rotation. We
obtained the tunneling probability for the incoming and outgoing
particles and correctly recover the Hawking temperature. We have
explicitly calculated the action as well. An indication of the
generality of our results is that, in appropriate limits, they
reduce to those for the Kerr and Schwarzschild black holes.

\acknowledgments

We are thankful to Douglas Singleton for some useful comments.

\end{document}